\documentclass[a4paper,11pt]{article} \usepackage{pos} \usepackage{subcaption}

\title{Reproducibility and Open Science in Lattice Quantum Field Theory}

\author[a]{Ed Bennett}
\affiliation[a]{Swansea Academy of Advanced Computing, Swansea University, Bay
  Campus, Fabian Way, Swansea SA1 8EN, United Kingdom}
\emailAdd{e.j.bennett@swansea.ac.uk}

\ExplSyntaxOn
\seq_put_right:Nn \l_authors_for_head_s{Ed\ Bennett\ et\ al.}
\ExplSyntaxOff

\author[b]{Andreas Athenodorou}
\affiliation[b]{Computation-based Science and Technology Research Center, The Cyprus Institute, 20 Kavafi Str., Nicosia 2121, Cyprus}
\emailAdd{a.athenodorou@cyi.ac.cy}

\author[c]{Louise Chisholm}
\affiliation[c]{Advanced Research Computing Centre, University College London, Gower Street, London WC1E 6BT}
\emailAdd{l.chisholm@ucl.ac.uk}

\author[d]{Anna Hasenfratz}
\affiliation[d]{Department of Physics, University of Colorado, Boulder, Colorado 80309, USA}
\emailAdd{anna.hasenfratz@colorado.edu}

\author[e]{Carsten Urbach}
\affiliation[e]{Helmholtz-Institut f\"ur Strahlen- und Kernphysik
  (Theorie) and Bethe Center for Theoretical Physics, Universit\"at Bonn, 53115 Bonn, Germany}
\emailAdd{urbach@hiskp.uni-bonn.de}

\abstract{
  Reproducibility and Open Science are increasingly discussed as essential aspects of the research process.
  While there are areas where the Lattice community has been ahead of the curve
  with respect to the broader research world in this space,
  including early adoption of open publications via the arXiv,
  and the introduction of the International Lattice Data Grid in the 2000s,
  there are other areas
  where lattice practitioners could benefit from practices already adopted in other disciplines.
  In this Contribution,
  we report the outcomes of a panel discussion on this topic at the Lattice 2024 conference;
  after a discussion on motivations for work in this space,
  and introductory discussions of the relevant experiences of the panelists,
  we provide summaries of answers to the questions posed by the audience in the panel.
}

\FullConference{The 41st International Symposium on Lattice Field Theory (LATTICE2024)\\
 28 July - 3 August 2024\\
Liverpool, UK\\}

\begin{document}
\maketitle

\section{Introduction}

The Turing Way project defines a result as \emph{reproducible}
if when the same analysis steps are performed on the same dataset,
the same result is consistently produced~\cite{the_turing_way_community_2023_7625728}.
One may consider this to be trivial,
or that any result not meeting this criterion could not be considered to be a meaningful scientific output;
however,
a surprising amount of research does not pass this low bar.
In particular,
work that does not share the data it analyses,
and the software workflow used to perform this analysis,
is almost never fully reproducible:
the narrative of a paper cannot fully specify every analysis step performed,
and unless an analysis is fully automated,
human error means that even a fully-specified analysis is
unlikely to exactly reflect the analysis that was ultimately performed.

Open Science is
the movement to make all outputs of research,
as well as all the elements used throughout the lifetime of the research project,
accessible to all of society.
This includes not just papers,
but also data, software, experimental samples, access on physical infrastructure
and other outputs too diverse to list.
The goal is not only to share with trusted colleagues,
but with anyone who may benefit from access.
It goes hand-in-hand with reproducibility:
it is hard to enable reproducibility without being open,
and being open encourages reproducibility,
and allows others to verify it.
It is however always a goal to be strived toward
rather than an achievable end-point;
there will always be limits on what is feasible to share---for example,
making a petabyte of field configurations instantly accessible to anyone on the planet
is prohibitively expensive,
and likely not productive.

In this contribution,
we summarise the panel discussion held during the Lattice 2024 conference.
After defining the relevant terms and further motivating the discussion,
we provide brief updates on topics relevant to the panellists' experience
likely to be useful and interesting to a broad audience of lattice theorists,
and then address questions presented to the panel,
which form the majority of the contribution.

\section{Definitions and Motivation}

Related to the concept of reproducibility
are the ideas of replicability and robustness,
which cover more stringent requirements than the definition of reproducibility above,
while still covering topics that some would put under that umbrella.
To quote The Turing Way project,
``A result is replicable
when the same analysis performed on different datasets produces qualitatively similar answers,''
and
``A result is robust
when the same dataset is subjected to different analysis workflows to answer the same research question \dots
and a qualitatively similar or identical answer is produced.''
Both would frequently be considered to be
essential requirements for a work to be considered useful computational science.

Another term that is frequently used in discussions of Open Science is FAIR,
the idea that data (and software) should be
``Findable, Accessible, Interoperable, and Reusable''~\cite{wilkinson2016fair}.
Data are not useful if they cannot be found,
so they should be made easily discoverable both by humans and by machines,
by using persistent identifiers, rich metadata, and searchable indexes.
One found,
mechanisms to access them should be standardised an open,
even if authorisation and authentication are needed,
and metadata should remain accessible even if data become unavailable.
To enable data from different sources to be easily used in the same analysis,
data and metadata should make use of standardised formats
and definitions that are themselves FAIR\@.
And to enable data to be reused,
they should be made available with explicit permission for this,
and carry sufficient information to trace their provenance.

There are many good reasons to support reproducibility and Open Science,
of which we will outline four,
ordered from idealistic to pragmatic.

For over three and a half centuries,
the motto of the Royal Society,
one of the most influential organisations on the modern scientific method,
has been
\emph{nullius in verba}:
``don't take anyone's word for it''.
Science is built on the principle that
one should be able to reproduce and replicate the work of others
before it is trusted.
Computational science should be no different:
it is incumbent on us to show our work,
and not treat the computer as a black box that dispenses wisdom.
While many lattice HPC codes are public,
most work either does not specify the software used or uses unpublished modifications,
and very little work also includes the software workflow used for post-HPC data analysis~\cite{Bennett:2022klt}.
Further,
a majority of respondents to a 2022 survey indicated that they were
``not very'' or ``not at all'' familiar with the concepts of
FAIR data and persistent identifiers~\cite{Athenodorou:2022ixd},
indicating that there is still work to be done to get the community to the point where
it has the skills to make its work open and reproducible.

The vast majority of research in lattice quantum field theory
is funded by governments,
which raise revenue by taxing the broader public.
Since the public are ultimately funding the effort,
it stands to reason that they should also benefit from the results.
This is the argument that has led to the growth of open access publications,
building in many ways on the success of the arXiv~\cite{arXiv};
while in lattice,
one has been able to read virtually any paper in the field free of charge for many decades
(at least in preprint form),
and other disciplines are now beginning to catch up on this,
the same has not been true for data.

Relatedly,
lattice quantum field theory computations form a primary use of Tier 0 HPC usage worldwide.
This is resource-intensive both in production and operation,
and as such
significant amounts of energy consumption and environmental impact occur during
the production of configurations,
the analysis of primary data,
and the generation of secondary data such as two- and three-point functions.
This carries a substantial carbon footprint,
whcih is increasingly at odds with
the global push for sustainability and minimal environmental impact.
Reproducing data that have already been generated
by another lattice practitioner or research group
is energetically inefficient,
duplicating both time and environmental costs unnecessarily.
By prioritising open and reproducible science,
we can reduce redundant computations,
conserve energy,
and work towards a more sustainable approach to large-scale computational research.

This leads into the third,
and most pragmatic,
reason for moving towards reproducible, Open Science:
our funders are beginning to require it.
The Science and Technology Facilities Council (STFC),
which funds particle physics research in the United Kingdom,
has in recent years specified in its Scientific Data Policy that
``Data resulting from publicly funded research should be made publicly available\dots
unless there are specific reasons
(e.g.\ legislation, ethical, privacy and security)
why this should not happen''~\cite{stfc-sdp}.

\section{Context}

In this section,
we share information drawn from the panellists' experience
that is relevant and provides context to the subsequent discussion.


\subsection{Current experiences of data sharing and reproducibility verification}


In Lattice QCD,
it is common to compute observables from first principles yet to be seen in nature,
as predictions to be tested at collider experiments.
We also compute quantities that have been observed previously,
but at finer resolution,
to better test the precision of the Standard Model.

In Lattice Quantum Field Theory Beyond the Standard Model,
it is common to probe unconventional systems,
such as those with conformal or exotic phases in the ultraviolet or infrared,
and to compute unconventional quantities,
such as order parameters of exotic phases,
scaling dimensions,
and the renormalisation group beta function.

In all cases,
where controversies or tensions arise between different groups' results,
it is vital to be able to settle them efficiently.
Ideally,
this should include subjecting all data from different groups
to the same analysis process.
To enable this,
we must share our data
and the code used to analyse it.

\begin{figure}
  \includegraphics[height=5cm]{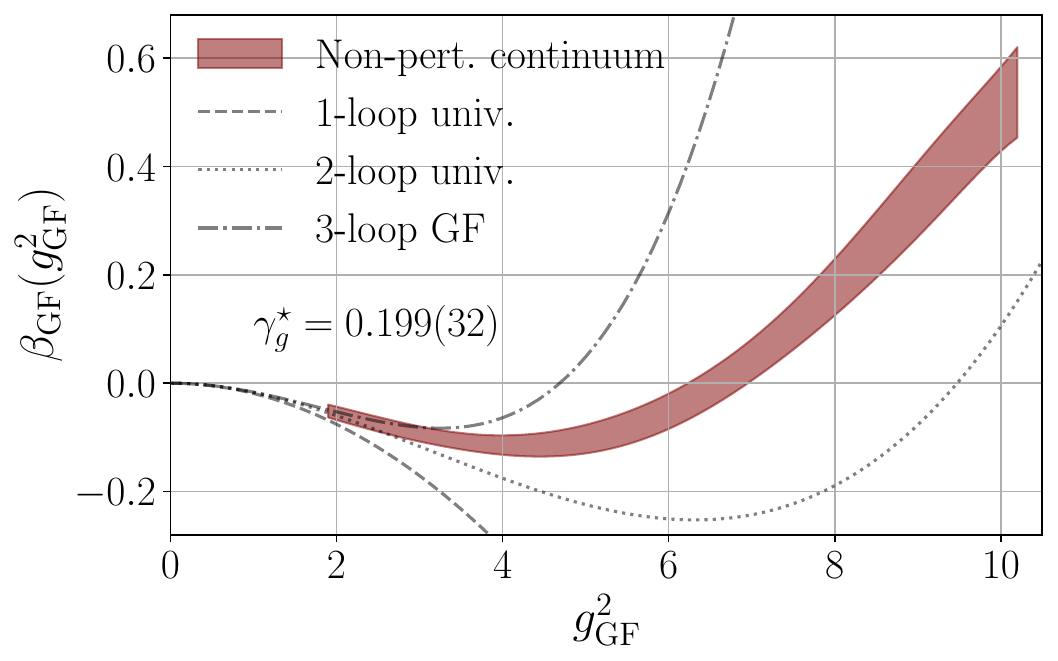}
  \hfill
  \includegraphics[height=5cm]{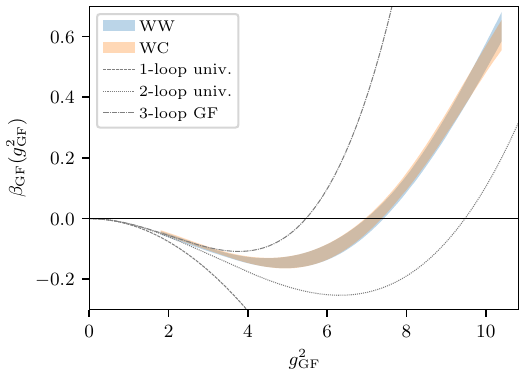}

  \caption{\label{fig:gfbeta}
    Left:
    The $\beta$ function of the SU(3) theory with twelve fermion flavours transforming in the fundamental representation,
    using ensembles generated with additional smeared Pauli--Villars fields,
    computed from data for the gradient flow of the Wilson action~\cite{peterson_2024_10719052},
    as presented in Ref.~\cite{Hasenfratz:2024fad}.
    Right:
    The same data~\cite{peterson_2024_10719052},
    analysed using an alternative workflow~\cite{bennett_2024_13362605},
    and presented in Ref.~\cite{Bennett:2024qik}.
  }
\end{figure}

While a 2022 survey~\cite{Bennett:2022klt} showed
few lattice publications share their data,
this has showed signs of changing.
For example,
since 2022,
the TELOS Collaboration~\cite{telos} has shared data and analysis workflows
for all of its journal publications presenting new data.
This year,
A.H.~\emph{et al} shared the data~\cite{peterson_2024_10719052} underlying Ref.~\cite{Hasenfratz:2024fad};
E.B.~was then able to write an independent analysis workflow~\cite{bennett_2024_13362605}
that qualitatively verified the results shown in Ref.~~\cite{Hasenfratz:2024fad}
(as compared in Fig.~\ref{fig:gfbeta}),
before using the same workflow components in the context of a
different theory~\cite{Bennett:2024qik}.
(Other recent work from members of the panel that are reproducible using available code and data include
Refs.~\cite{Ostmeyer:2024qgu,Athenodorou:2024rba,Bennett:2024tex,Bennett:2024wda,Bennett:2024cqv}.)
In this way,
sharing openly accelerates the progress of science,
by allowing others to more rapidly adopt techniques that we develop,
and trust in the results that they give.

\subsection{The International Lattice Data Grid}
\label{sec:ILDG}


The International Lattice Data Grid
(ILDG)
project was first announced at LATTICE 2002~\cite{Davies:2002mu},
and the first metadata schema,
QCDml,
was presented at LATTICE 2004~\cite{Maynard:2004wg}.
This places Lattice as a very early mover in the space of open, FAIR data,
over a decade before the term FAIR was first coined~\cite{wilkinson2016fair}.
The federated grid infrastructure
means that the ILDG organisation provides few resources directly;
it defines metadata schemas,
binary formats,
and standards for metadata and file catalogues,
and provides a Virtual Organisation
(VO)
that can be used to provide common authentication.
These may be deployed by Regional Grids,
which may then provide a standard interface,
rather than each collaboration providing their own infrastructure with different conventions,
which a prospective publisher or consumer of data
would need to learn to use from scratch
each time they work with a new collaboration.

ILDG-compatible infrastructure enables FAIR sharing of lattice data,
both within collaborations
(including internal data that have not yet been made public),
and across the community
(increasing the reach and impact of each dataset,
and maximising the amount of scientific value obtained
for each unit of public money, energy, or clock cycle).
It supports good scientific practice,
helping to enable reproducibility
both by better documenting the generation process for ensembles,
and allowing others to reproduce observables computed based on them.
It is also invaluable in being able to demonstrate compliance
with requirements from funders to manage and share data appropriately.

Recently, there has been significant progress made in modernising ILDG,
including defining requirements for ILDG 2.0~\cite{DiRenzo:2024lgo},
replacement of the registration and authetication mechanisms, and
re-implementations of the metadata and file catalogues. 
Significant progress towards delivering these has been reported
at this conference~\cite{ILDG20:2025}.
The work done and community built around it
are significant assets to the field,
and provide a platform that can be further built on.

To date,
ILDG primarily targets gauge field configurations;
however,
there is scope to build on this to allow
better interoperability of other kinds of lattice data, like for
instance bootstrap samples of derived quantities,
rather than each group or software project
needing to define its own data structures and metadata schemas
for the same classes of data. There is also ongoing work to attach
digital object identifiers (DOIs) to data sets and make them citable.

This progress must of course respond to the needs and requirements of the community,
and avoid pursuing activity for its own sake.
Effort will be needed from across the community
to make this a success;
this may include leveraging
new national funding schemes for research data management;
this has recently included the PUNCH4NFDI
(\textbf{P}articles,
\textbf{U}niverse,
\textbf{N}u\textbf{C}lei and
\textbf{H}adrons
\textbf{for} the
\textbf{N}ationale
\textbf{F}orschungs-\textbf{D}aten
\textbf{I}nfrastruktur---National Research Data Infrastructure%
~\cite{punch4nfdi}
)
programme in Germany,
which funded significant amounts of development time towards ILDG 2.0.

\subsection{European Union initiatives}
\label{sec:EU}

\subsubsection{The Open Science Policy in the European Union}
\label{sec:European_Policy}

Materializing Open Science is a key focus of European research policy. Various policies, initiatives, and frameworks are being created and executed to make European science and research more open, enhancing their efficiency, productivity, transparency, and robustness, while also ensuring they meet the needs and expectations of both policymakers and society.

The goal is to position Europe as a leader in harnessing the benefits of Open Science by fostering a more open research system, encouraging collaboration among scientists and across different disciplines, and engaging with society as a whole. Open Science promotes sharing and cooperation, speeding up discoveries, enhancing research quality, and making science more influential and relevant to human and societal progress.

Achieving this requires dismantling barriers, creating incentives, and supporting data-intensive research and extensive knowledge sharing, as well as encouraging a scientific perspective in public discourse. Adopting Open Science as the standard approach to research involves enhancing the broad array of practices that constitute Open Science and developing the necessary tools and conditions to support it.

The new and ongoing EU Framework Programme for Research \& Innovation, \emph{Horizon Europe},
integrates an approach where Open Science is the ``modus operandi.''
In the previous EU Framework,
Horizon 2020,
Open Science was largely limited to open access.
However,
there is now a strong shift in EU Framework Programmes
toward stricter implementation of comprehensive Open Science practices.

At a proposal stage Open Science is assessed under
both the ``Excellence'' and the ``Quality and Efficiency of Implementation'' criteria.
\emph{Horizon Europe} differentiates between mandatory and recommended Open Science practices.
However,
to achieve a high evaluation score,
it is essential to integrate not only the mandatory practices but also the recommended ones.
As the \emph{Horizon Europe} Programme Guide states,
``proposers are expected to adopt when possible and appropriate for their projects.''

The status of Open Science practices within \emph{Horizon Europe},
particularly relevant to the field of Lattice Gauge Theories, is as follows.

\begin{description}
  \item[Mandatory Open Access to Publications]
        Researchers must ensure open access to all peer-reviewed scientific publications resulting from their work.
        A machine-readable electronic copy of either the published version
        or the final peer-reviewed manuscript
        must be uploaded to a trusted scientific repository as soon as possible,
        but no later than the date of publication.
        Open access should be provided immediately upon publication.
  \item[Open Access to Research Data]
        Open access to research data should be provided ``as soon as possible.''
        Researchers are required to implement measures that ensure the reproducibility of research outputs.
        This includes making the necessary research data available under the principle
        ``as open as possible, as closed as necessary,''
        meaning data should be openly accessible unless there is a compelling reason to restrict access.
  \item[Licensing and Attribution]
        Publications should be released under the most recent version of the
        Creative Commons Attribution International Public License (CC BY)
        or a license with equivalent rights.
        This ensures that the work can be
        freely used,
        distributed,
        and reproduced,
        provided that appropriate credit is given to the authors.
  \item[Research Outputs and Validation]
        Authors are required to provide comprehensive information about
        the research outputs,
        tools,
        and instruments
        necessary to validate the conclusions of their scientific publications
        or to validate and re-use research data.
        This includes providing access to
        software,
        algorithms,
        models,
        workflows,
        and other outputs essential for validation.
  \item[Data Management and FAIR Principles]
        Researchers must develop and adhere to a Data Management Plan (DMP)
        to ensure the responsible management of research data and metadata.
        This plan should align with the FAIR principles
        (Findable, Accessible, Interoperable, Reusable)
        and cover
        all research outputs,
        including
        publications,
        data,
        software,
        algorithms,
        protocols,
        models,
        and workflows.
  \item[Documentation and Access for Validation]
        The research outcomes should be adequately documented in a Research Output Management Plan.
        Both digital and physical access to the results necessary
        to validate the conclusions of scientific publications should be ensured.
\end{description}

In summary, \emph{Horizon Europe} mandates that researchers in Lattice Gauge Theories, like all participants, adopt rigorous Open Science practices. This includes immediate open access to publications, responsible data management, and ensuring that all research outputs are accessible and reproducible in line with the highest standards of transparency and openness.

\subsubsection{The European Open Science Cloud (EOSC)}
\label{sec:EOSC}

To facilitate and practically support Open Science in the European region,
the European Union (EU) has initiated the European Open Science Cloud (EOSC).
EOSC aims to create a
federated,
open,
and multi-disciplinary environment
where European researchers,
innovators,
businesses,
and citizens can
publish,
discover, and reuse
data,
tools,
infrastructures and services
for research,
innovation,
and educational purposes.
The EOSC platform is expected to act as an one-stop shop,
in order to exemplify the life of the researcher.
This environment functions under clear conditions designed to ensure trust and protect the public interest.

EOSC facilitates a significant shift across scientific communities and research infrastructures towards,
a.~Seamless access,
b.~FAIR (Findability, Accessibility, Interoperability, and Reusability) management,
and c.~Reliable reuse of research data and all digital objects produced throughout the research life cycle
(e.g., methods, software, and publications).

The ultimate goal of EOSC is to create a
``Web of FAIR Data and Services''
for European science,
enabling the development of a wide range of value-added services.
These may include
data visualisation,
advanced analytics,
long-term information preservation,
and monitoring the adoption of Open Science practices.
For Lattice Gauge Theories,
one potential opportunity could be integrating ILDG as a service within EOSC,
leveraging core EOSC features such as
Authentication and Authorization Services,
the Help Desk,
and Monitoring tools.

The EOSC is acknowledged by the Council of the European Union as
one of the 20 actions in
the policy agenda for 2022-2024 of the European Research Area
(ERA),
specifically aimed at enhancing Open Science practices in Europe.
It is also identified as the
``science, research, and innovation data space,''
fully integrated with other sectoral data spaces defined in the European data strategy.

The implementation of the European Open Science Cloud (EOSC) has been an ongoing, long-term process of alignment and coordination, led by the European Commission since 2015. This initiative has engaged a wide range of stakeholders from across the European research community. In practical terms, the development of EOSC has been carried out through targeted European calls, initially launched under the Horizon 2020 framework, which focused on creating the platform, establishing governance structures, and building capacity. Among these initiatives was the NI4OS-Europe project, which aimed to enhance capacity building in the regions of Southeast Europe and the Eastern Mediterranean. In this context, AA had a role as a Work Package leader, particularly in the task of involving scientific communities, such as Computational Physics, in the EOSC.

The new European framework,
\emph{Horizon Europe},
strongly supports the development of innovative services to promote the implementation of Open Science,
including Workflow Orchestration Platforms
and capacity-building initiatives focused on specific scientific fields,
as well as the broader ``long tail'' of science.
One notable example is the INFRAEOSC project,
OSCARS~\cite{oscars},
which brings together leading European Research Infrastructures (RIs)
to advance Open Science in Europe.
These RIs are organised into five Science Clusters,
including Nuclear and Particle Physics.

The primary goals of OSCARS are to support FAIR (Findable, Accessible, Interoperable, Reusable) data management, provide federated thematic services, and ensure harmonised access to data, tools, and training. OSCARS operates through Open Calls for expressions of interest to develop Open Science projects. Lattice Gauge Theories, with their unique research focus, align well with OSCARS' objectives and could actively participate in projects supporting Open Science in this specialised field.

\subsubsection{Europen e-Infrastructures and the EOSC}
\label{sec:e-Infras}

A successful implementation of Open Science principles requires the availability of tools and infrastructures that empower researchers and stakeholders to adopt and practice these principles effectively. Alongside the European Open Science Cloud (EOSC), several e-Infrastructure initiatives have been instrumental in enabling the realization of Open Science. Recently, an assembly of key e-Infrastructures has been established to further this mission, providing a coordinated approach to supporting Open Science. This assembly comprises GÉANT, OpenAIRE, EUDAT, EGI, and PRACE. A brief overview of these e-Infrastructures is provided below.

\begin{description}
    
\item[GÉANT]
  (Gigabit European Academic Network)~\cite{geant}
  is Europe’s leading collaboration for networking, e-infrastructure, and related services,
  dedicated to advancing research and education across the continent.
  The organization develops, delivers, and promotes state-of-the-art network infrastructure and services,
  fostering innovation and facilitating knowledge exchange
  among its members, partners, and the broader research and education community.
  GÉANT provides high-bandwidth connectivity across Europe,
  supporting major scientific endeavors such as High-Energy Experimental Physics,
  as well as widely-used services like eduroam,
  which enables seamless roaming for researchers and students.

\item[OpenAIRE]
  (Open Access Infrastructure for Research in Europe)~\cite{openaire},
  established in 2018 as the non-profit legal entity OpenAIRE A.M.K.E.,
  is committed to fostering sustainable open scholarly communication infrastructures
  and promoting research excellence across Europe and beyond.
  Through collaborative efforts and proactive initiatives,
  OpenAIRE delivers innovative services,
  shapes influential policies,
  and provides comprehensive training to support the adoption of Open Science.
  Since its inception in 2009,
  OpenAIRE has played a pivotal role in the development and advancement of the European Open Science Cloud (EOSC).
  Among its many services,
  OpenAIRE provides Zenodo for publishing research outputs
  and Argos for creating Data Management Plans,
  alongside numerous other tools designed to empower researchers and institutions.

\item[EUDAT]
  Collaborative Data Infrastructure (EUDAT CDI)~\cite{eudat},
  is one of Europe’s largest ecosystems of integrated data services and resources designed to support research.
  Backed by a network of over 20 European research organizations and data and computing centers,
  EUDAT offers comprehensive solutions for managing the entire research data lifecycle.
  Its services include B2DROP for syncing and sharing active data,
  B2SHARE for storing and publishing stable datasets,
  and B2SAFE for ensuring long-term data preservation through replication.
  Additionally,
  researchers can register data with B2HANDLE and enjoy seamless access to all services using their institutional credentials via B2ACCESS.

\item[The EGI Federation]
  (European Grid Infrastructure)~\cite{egi}
  offers a flexible and scalable digital research infrastructure
  that empowers tens of thousands of researchers across diverse scientific disciplines.
  Through its collaborative network of hundreds of public and private service providers,
  EGI provides access to advanced computing and data analytics resources sourced from across Europe and beyond.
  The Federation’s services include
  distributed high-throughput computing,
  cloud solutions,
  efficient storage and data management,
  co-development of innovative technologies,
  expert consultation,
  and comprehensive training opportunities.

\item[PRACE]
  (Partnership for Advanced Computing in Europe)~\cite{prace}
  is committed to advancing the interests and addressing the needs of HPC users and related technologies,
  including Artificial Intelligence,
  Quantum Computing,
  Cloud Computing,
  and Data Science,
  across Europe.
  As a collective of HPC users and centers,
  PRACE fosters high-impact research and innovation in a wide range of scientific and industrial fields,
  ultimately enhancing Europe’s scientific and technological capabilities.

\end{description}

The newly formed assembly of these five e-Infrastructures is committed to promote Open Science Principles.
Namely, the assembly advocates for transparency, accessibility, and reusability of research outputs.
In addition,
it aligns its activities with Action 1 of the ERA Policy Agenda,
which focuses on enabling the open sharing and reuse of research outputs.
The assembly is also committed to supporting research communities in
integrating their data into the European Research Area and the web of FAIR data.
Furthermore,
it aims to provide services and strategies to make data management consistent with Open Science practices. 

The assembly plays a crucial role in supporting and advancing the goals of the European Open Science Cloud (EOSC). As a collaborative initiative, the assembly aligns its activities with EOSC’s vision to create a seamless, pan-European infrastructure for Open Science. Through its member e-Infrastructures, the assembly contributes to EOSC by providing the necessary digital infrastructure, data services, and research resources that enable open access, data sharing, and collaboration across Europe. By fostering coordination and alignment, the assembly ensures that the e-Infrastructures complement and enhance EOSC, working together to strengthen the overall ecosystem for Open Science, increase interoperability, and support the efficient management and dissemination of research data.

\subsection{The Square Kilometre Array (SKA)}
\label{sec:the_square_klilometer_array}

The Square Kilometre Array is
a next-generation radio astronomy facility,
currently under construction.
The low-frequency and mid-frequency antennae are being built
in Australia and South Africa respectively,
with the infrastructure to combine the observations based at
the Jodrell Bank Observatory in the United Kingdom.

SKA is expected to produce tens of petabytes per year of data.
The current accepted practice in observational astronomy
is that after a period of exclusivity
(also known as a ``proprietary period'' or ``embargo'')
during which the group who designed the study
may prepare their work based on it without being ``scooped'',
all observation data are made public,
so that other groups may benefit from phenomena observed
that were not the target of the original work.

Both the collection of the raw data from the instruments,
and the retention and distribution of processed data openly,
require significant infrastructure
due to their large size
and the distributed nature of the facility.
The SKA project is in the process of developing
an end-to-end system from telescope time proposal to science delivery,
that will also enable FAIR data sharing,
including supporting appropriate metadata standards,
sharing policies,
and proprietary periods.

\subsection{Challenges, Opportunities, and the FAIR Data Accelerator}

There are a number of barriers to open sharing of data and software;
while some of them are technical
(for example,
availability of appropriate metadata standards and hardware platforms),
others are social or cultural.
Researchers must be willing to make their data open,
and have the skills to do so.
The FAIR Data Accelerator Pilot~\cite{fair-data-pilot}
is a project that aims to understand what these cultural barriers are,
and develop strategies to overcome them.
Initial work~\cite{fierro2024} indicates that there are two key themes:
epistemic uncertainties
(fears,
concerns,
and vulnerabilities),
specifically,
the fear of judgement,
and the fear of losing control of data;
and cultural reluctance
(practices,
structures,
and policies),
including the perception that sharing is not a priority,
that there is a lack of recognition or something in exchange,
and that leaders do not engage.
Overcoming these barriers will create a sense of epistemic trust,
allowing researchers to better buy in to the process.

However,
there are also new opportunities in this space.
There is room for new approaches to proposing and undertaking research,
building on work released openly by others for rapid results,
rather than needing to start from scratch each time.
Science is increasingly a team endeavour,
and there is the opportunity for new professions to fill skill gaps in teams,
such as Data Stewards and Research Software Engineers.
There are new avenues for funding,
to fund infrastructure and professionals focusing on open sharing,
as well as work harnessing existing data resources.

One particular challenge appears to be the additional workforce
required to make data openly available: data curation, description and
documentation takes a significant amount of time. This makes it even
more important to join forces at the international level to address
this challenge, but it will likely not be feasible to do so without
dedicated additional human capacity.

\section{Questions and Answers}

In this section,
we summarise the questions asked of the panel,
and the responses given by the panel,
as well as providing more detail or context in some cases.

\subsection{How much new stuff do you have to learn? What effort is needed, and what benefits do you get back?}

Improving reproducibility and openness is open-ended;
there is no lower bound,
and all effort is likely to lead to a measurable improvement.
That said,
as is frequently the case,
the effort put in is positively correlated with the return.

Learning to take the code that you already use,
and the data that you are already generating,
and share them using Zenodo
would be one starting point,
as would remembering to precisely specify software information
in the same way that we already do for grants and HPC time.
This requires a little extra time at publication,
but already gives a measurable improvement
in others' ability to understand and reproduce your work.

Building the analysis workflow for your next work
such that it is reproducible by design
is somewhat more work,
and requires more thought about the structure of the code
than doing each step by hand;
it also benefits from learning the basics of
technologies such as workflow managers.
However,
the benefits are significant later in the project,
where the time required to make changes as new data arrive,
or as tweaks are needed during the publication process,
is greatly reduced.
Additionally,
when it comes time for the next project;
modular,
reproducible workflows are
much easier to repurpose and recycle for future work,
allowing successive projects to move forward with greater velocity.

Potentially the hardest task is
to take an analysis that is already in progress or near-complete,
and retroactively automate it.
The benefit here is that it will expose numerical errors in the original analysis,
which otherwise would have made it into a publication.
E.B.~has done this exercise for half a dozen publications,
with multiple collaborations,
and in every case has identified and rectified inconsistencies
either between the data and the text,
or between different presentations of the same data in the same work.


\subsection{Is it possible to make source code open access but remain competitive?}

This question could be reversed:
as Open Science becomes more common,
will it be possible to keep source code closed and remain competitive?
If sharing source code openly is the norm,
then reviewers and readers might reasonably question
what a collaboration is trying to hide if it doesn't share its code,
and lend more credulity to groups that do.

Looking outside of academia,
open source software has not only proven competitive,
it has dominated many areas.
Even proprietary software now makes heavy use of open source libraries,
as can be seen in the Acknowledgements section of such software's documentation.
(Apple's macOS,
for example,
attributes over 200 such open-source software packages.)
Further,
maintainers of such packages are frequently recruited to work for
the technology companies making use of them,
since they are world leading experts in deploying and enhancing them.
The same can apply in academia:
a powerful way to bring expertise on a particular technique into a group
is to recruit the developer of an open implementation of that technique.
Being able to see the code and documentation they have written
gives more insight into their capabilities than any interview could.

\subsection{Where does the money come from to pay for data repositories, data stewards, research software engineers? Does it require sacrificing person-effort on research?}

In many regions there are funding schemes specifically targeting this kind of work;
any funding from such routes would obviously not be at the expense of more general research funding.

However,
as research moves more in the direction of needing openness and demonstrable reproducibility,
data stewardship and research software engineering tasks
increasingly become a part of the basic requirements of performing research.
At this point,
resourcing specific roles who can perform such tasks more quickly because it is their bread and butter
can deliver a net gain in scientific throughput
compared to each researcher needing to fend for themselves.
While it is too early to say definitively,
one would hope that such a change would enable more science to be done per unit funding,
rather than less;
a healthy ecosystem of
open data,
software,
and workflows
should enable researchers to reach results more quickly
compared with always starting from scratch.

\subsection{Storage of correlation functions, etc.\ often takes $\sim$10 GB or more and must be maintained indefinitely. What is a good place/site to store these data?}

There are no organisations that can realistically commit to maintaining storage indefinitely.
However,
most requirements for data retention are usually of the order 10--20 years;
storage beyond this point may be an aspiration,
but is not considered reasonable to request up front.

Zenodo~\cite{zenodo}
is a service hosted by CERN\@.
It commits to retain data ``for the lifetime of the repository.
This is currently the lifetime of the host laboratory CERN,
which currently has an experimental programme defined for the next 20 years at least''~\cite{zenodo-policies}.
It provides for free upload of records up to 50GB in filesize
(where typically one record would be expected to correspond to the data for a single paper),
and provides a persistent identifier for each (version of each) record
in the form of a Digital Object Identifier (DOI),
which can be cited in other work.

Alternative services may be provided by individual institutions;
for example,
the University of Bielefeld's Unibi service
has been used to retain and share data from multiple lattice publications~\cite{Bennett:2022klt}.
Another alternative in the EU is the B2SHARE service offered by EUDAT~\cite{eudat}.


\subsection{Making software user-friendly is a lot of work. Why should we sacrifice research time for it?}

While having user-friendly software is usually welcome,
it is worth first recognising that it is not essential for reproducibility.
The first step to reproducible, open research
is to publish the exact code that was used to generate the outputs published in a paper,
regardless of usability.
This allows others looking to replicate or generalise the work
to identify specific implementation details that would likely be skipped in the narrative;
for example,
choice of number of bootstrap samples,
or particular normalisation conventions.
While this still requires a non-zero amount of time,
it is a significantly smaller request to make of someone's time
than polishing a piece of code into something immediately usable by others.
(The justification for spending this smaller time
is likely the same as why we write research articles at all:
we want others to understand the work we have done,
to contribute to the scientific body of knowledge,
and to gain recognition for the work done,
including via citations.)

Most scientific code exists on a spectrum from
scripts that will be used once and never revisited
(for example,
to generate a plot comparing two fits,
which was only ever included in an internal collaboration note)
to software that has become foundational to work across many disciplines
(for example,
IPython~\cite{PER-GRA:2007},
developed by former lattice field theorist and A.H.'s former student Fernando Peréz).
When deciding on the level of effort to place into making software user-friendly,
one must take this into account.
Another consideration is the amount of future effort saved---for example,
will polishing a piece of code now
reduce the amount of time you will need to spend teaching future students to use it,
or make it quicker to get started the next time you need to use it?

\subsection{How can we effectively convey to hiring and promotion committees that the work that goes into openness, FAIR data, and reproducibility should count as physics?}\label{sec:recognition}

Changes of this sort cannot happen without being championed by those with influence.
Many of those present in person for this panel,
or who are reading this contribution,
sit on such committees,
and have influence on developing their terms of reference.
We would refer such readers to,
for example,
Refs.~\cite{puebla2024ten,ukrnguide}
for recommendations on how they can best use their influence
to drive progress in this space.

For all of us,
even those without such influence,
normalising the idea of \emph{nullius in verba}
will drive greater understanding that
the work to enable this is essential.
Where data and/or code are promised ``on request'',
in a work that you are seeking to build from,
do not hesitate to request them.
When peer reviewing work where this commitment is made,
it is especially important to verify that the authors
are able to deliver on it;
conversely if it is not made,
then it is valid to ask why data are not provided,
and why the work should be trusted in their absence.


\subsection{I want to start engaging with reproducible Open Science; what would be a good first step? Are there resources available, or networks that I can join?}


Introductory training materials on Reproducibility and Open Science
in the context of lattice field theory
are available through the Lattice Virtual Academy (LaVA)~\cite{lava}.
These will continue to be developed.
Other,
more general,
resources that could be a suitable starting point include
The Turing Way,
which is handbook for
reproducible,
ethical
and collaborative data science~\cite{the_turing_way_community_2023_7625728}
and NASA's Transform to Open Science (TOPS) course~\cite{nasa-tops}.

The topic is also relevant to the interests of the ILDG,
who welcome new members on their mailing lists.

\subsection{What is a realistic embargo time? For configurations? For other data?}

Many large project impose an embargo on data,
where the data supporting one or more publications
are not made available outside the collaboration
for some period of time.
To meaningfully define a realistic embargo,
we must define when such an embargo would start,
as well as understanding why such embargoes are used.

For example,
in the astronomy community,
observations from telescopes can be  embargoed for up to 12--18 months after the data are collected~\cite{skaafrica-call}. 
This allows the
(potentially very small)
team who were awarded time on the telescope
to analyze and publish
before the data are publicly released,
without being ``scooped''.
This approach also contributes to
equality, diversity and inclusion efforts
supporting researchers and students who
have access to limited computational resources,
heavy teaching loads
or other responsibilities.

However, the embargo period is not universally applied.
For example, The James Webb Space Telescope's policies 
only give smaller experiments an Exclusive Access Period
(with these still being encouraged to reduce or remove it),
with larger experiments receiving none by default~\cite{jwst-rights,jwst-open}. 

For lattice configurations,
which are frequently generated over a longer period of time,
and where groups may need to be conservative about releasing results with too few statistics,
it may be better to count any embargo time from first publication of results.
As in other communities,
there is always a need to balance the need of the community
to get the most value out of the shared resources committed to the work,
and the need of the originating research group
to achieve all of their stated aims with the data
without others beating them to the punch.

For smaller,
downstream data
such as correlation functions,
it is harder to imagine strong reasons
to not publish the data concurrently with the paper presenting their analysis.

We would not expect infrastructure providers such as ILDG
to dictate policies around embargo lengths;
rather,
they should facilitate whatever embargo is agreed upon.
The length should come from a conversation between
the research group,
the organisation funding the research
(for example,
funding council or HPC centre),
and the wider community.

\subsection{We depend on high-quality software, both for HPC and for later data analysis. How do we ensure that Research Software Engineers are resourced to write this}

Firstly,
we must recognise the work that is already being resourced.
Most HPC centres now employ dedicated RSEs to optimise software running on their systems,
and these resources have resulted in significant effort towards
porting lattice software to new architectures.
Much work of this sort is not published in the literature
although for open-source software it is typically made available;
one published example is Ref.~\cite{Rantaharju:2018emq}.

By necessity,
this work is almost always project-oriented,
focusing on one specific optimisation,
platform,
or feature;
for example,
porting to CUDA to support NVIDIA GPUs,
or optimising the performance of one specific Dirac operator implementation.
Conscientious RSEs will also implement
test suites and other documentation and refactoring
where it enables their work,
and contribute that back to the upstream code where it is welcome,
but this does not replace the need for long-term resourcing of code owners
and effort to do more low-level maintenance of software.

As of 2022,
physics and astronomy produced over 21\% of self-identified RSEs~\cite{Hettrick_International_RSE_Survey};
there is not a shortage of software engineering talent in our field.
However,
there needs to be recognition of software effort,
in order to retain this talent within the field.
With the support of funders,
this may be in the form of dedicated departmental or disciplinary RSE roles,
but there are other options as well.
As discussed in Section~\ref{sec:recognition} above,
recognition of software contributions and relevant metrics
in hiring and promotion practices
would avoid driving those who could make positive software contributions
to instead focus on applying the software in research,
or to leave the field in search of an occupation that better supports software effort.
Conversely,
as well as driving the reward structures to recognise software contributions,
we may also adjust software contributions to fit the reward structures.
Publishing research software as peer-reviewed articles is already possible in some contexts;
for example,
the Journal of Open Source Software~\cite{joss}
allows one to submit a short description of a software repository,
obtain peer review on the software,
and then obtain a citable DOI for the software and its description.
There may also be space for a dedicated journal
focusing on technical contributions to the field,
that do not directly generate new science
(as is required to publish in most journals in the field);
this could include software contributions,
algorithmic developments,
and even negative results
(which are widely underreported~\cite{fanelli2012negative,mlinaric2017dealing}).

\section{Conclusions}

The lattice community has had many early achievements in Reproducibility and Open Science,
including early adoption of the arXiv for open access publication sharing,
and development of the ILDG for sharing field configurations.
However,
there remain significant opportunities to do better,
including sharing field configurations more consistently,
providing downstream data products such as correlation functions to enable reproducibility,
and publishing data analysis workflows and software
so that techniques can more easily be tested and applied more widely.

We would encourage those interested in driving such work forward
to contact us,
and to join relevant mailing lists of the ILDG~\cite{ildg-organization},
where there may be scope to start one or more new working groups
focusing specifically on the topics discussed here.

\acknowledgments

A.A. was supported by
the Horizon 2020 European research infrastructures programme ``NI4OS-Europe”
with grant agreement no.\ 85764, by the Computation-based Science and Technology Research Center (CaSToRC) at The Cyprus Institute and by the ``EuroCC'' project
funded by the
``Deputy Ministry of Research, Innovation and Digital Policy and the Cyprus Research and Innovation Foundation''
as well as by the European High-Performance Computing Joint Undertaking (JU)
under grant agreement No.~101101903.

The work of E.B. has been supported by
the UKRI Science and Technology Facilities Council (STFC)
Research Software Engineering Fellowship EP/V052489/1,
by the EPSRC ExCALIBUR programme ExaTEPP (project EP/X017168/1),
and by the STFC Consolidated Grant No.\ ST/T000813/1.

L.C. acknowledges  support from UKRI STFC for the UK SKA Regional Centre (ST/X00256X/1) and the DSIT/UKRI Research Data Cloud Pilot Commission to DRIC: FAIR Data Accelerator pilot (ST/Z000505/1).

A.H. acknowledges support from DOE
grant DE-SC0010005.

The work of C.U. has been supported by the Deut\-sche
Forschungsgemeinschaft (DFG, German Research Foundation) as 
part of the CRC 1639 NuMeriQS – project no. 511713970.

\paragraph*{Open access statement}
For the purpose of open access, the authors have applied a Creative Commons
Attribution (CC BY) licence to any Author Accepted Manuscript version arising.

\paragraph*{Research Data Access Statement}
No new data were generated during the preparation of this work.

\bibliography{references} \bibliographystyle{apsrev}
\end{document}